\documentclass[
aps,
prb,
reprint,
showkeys,
twocolumn,
nofootinbib,
superscriptaddress,
]{revtex4-2}

\usepackage{natbib}
\usepackage{amssymb}
\usepackage{color}
\usepackage{amsfonts}
\usepackage{textcomp}
\usepackage{amsmath}
\usepackage[charter]{mathdesign}
\usepackage{easyReview}
\usepackage{microtype} % Improve typographic layout. TS
\usepackage{upgreek}
\usepackage{wasysym}
\newcommand{\ped}[1]{\ensuremath{_{\rm #1}}}
\newcommand{\apex}[1]{\ensuremath{^{\rm #1}}}
\definecolor{link}{RGB}{57,106,177}
\definecolor{darkgreen}{RGB}{0,128,0}
\usepackage{float}
\usepackage{graphicx}
\usepackage{hyperref}
\hypersetup{
	colorlinks=true,
	linkcolor=link,
	citecolor=link,
	urlcolor=link
}

\begin{document}

\title{Unusually weak irradiation effects in anisotropic iron-based superconductor RbCa$_2$Fe$_4$As$_4$F$_2$}

\author{Daniele Torsello}
\email{daniele.torsello@polito.it}
\affiliation{Department of Applied Science and Technology, Politecnico di Torino, Torino, Italy}
\author{Erik Piatti}
\affiliation{Department of Applied Science and Technology, Politecnico di Torino, Torino, Italy}
\author{Michela Fracasso}
\author{Roberto Gerbaldo}
\author{Laura Gozzelino}
\affiliation{Department of Applied Science and Technology, Politecnico di Torino, Torino, Italy}
\affiliation{Istituto Nazionale di Fisica Nucleare, Sezione di Torino, Torino, Italy}
\author{Xiaolei Yi}
\affiliation{School of Physics, Southeast University, Nanjing, China}
\author{Xiangzhuo Xing}
\affiliation{School of Physics and Physical Engineering, Qufu Normal University, Qufu, China}
\author{Zhixiang Shi}
\affiliation{School of Physics, Southeast University, Nanjing, China}
\author{Dario Daghero}
\affiliation{Department of Applied Science and Technology, Politecnico di Torino, Torino, Italy}
\author{Gianluca Ghigo}
\email{gianluca.ghigo@polito.it}
\affiliation{Department of Applied Science and Technology, Politecnico di Torino, Torino, Italy}
\affiliation{Istituto Nazionale di Fisica Nucleare, Sezione di Torino, Torino, Italy}

\begin{abstract}
We report on the effects of 3.5\,MeV proton irradiation in RbCa$_2$Fe$_4$As$_4$F$_2$, an iron-based superconductor with unusual properties in between those of the pnictides and of the cuprate high-temperature superconductors. We studied how structural disorder introduced by ion bombardment affects the critical temperature, superfluid density and gap values by combining a coplanar waveguide resonator technique, electric transport measurements and point-contact Andreev-reﬂection spectroscopy.
We find an unusually weak dependence of the superconducting properties on the amount of disorder in this material when compared to other iron-based superconductors under comparable irradiation conditions. The nodal multigap state exhibited by pristine RbCa$_2$Fe$_4$As$_4$F$_2$ is also robust against proton irradiation, with a two-band $d\mbox{-}d$ model being the one that best fits the experimental data.\\\\
Cite this article as: D. Torsello, E. Piatti, M. Fracasso, R. Gerbaldo, L. Gozzelino, X. Yi, X. Xing, Z. Shi, D. Daghero, and G. Ghigo. \href{https://doi.org/10.3389/fphy.2023.1336501}{\textit{Front. Phys.} \textbf{11}, 1336501 (2024)}.   
\end{abstract}


\maketitle

\section{Introduction}
Ion irradiation has proven to be an extremely useful tool for the investigation of fundamental properties of superconductors\,\cite{Korshunov,Efremov2011,Ghigo2018PRL}. In recent years, significant efforts have been devoted to the study of iron-based superconductors (IBSs)\,\cite{Ghigo2017scirep, Ghigo2020sust, Torsello2020PRAppl, Torsello2020sust}, hoping that understanding their pairing mechanism could lead to a comprehensive understanding of unconventional superconductivity\,\cite{Hirschfeld}. Although this high goal is still far from being achieved, several interesting findings were reported on these intriguing materials, contributing to deepen the knowledge of the effects of disorder on the superconducting condensate and related phenomena. In particular, besides the more standard result of enhancing the vortex pinning capability\,\cite{Torsello2020sust}, an irradiation-disorder-induced transition was observed between $s_\pm$ and $s_{++}$ states in two-band IBSs of the 122 family\,\cite{Ghigo2018PRL}, that contributed to confirm the $s_\pm$ state for the pristine compound, as predicted theoretically\,\cite{Korshunov}.

In the last few years, the attention moved to a newly discovered family of IBSs, the 12442 pnictide family\,\cite{Wang2016JACS, Wang2017SciChiMat, Wu2017prb}, that shows some peculiar properties resembling those of the cuprates\,\cite{Piatti2023LTP, Wang2020SciChi_pulsed, Yi2020NJP, Wang2019prb}. The 12442-type ACa$_2$Fe$_4$As$_4$F$_2$ (where A = K, Rb, Cs) compounds are obtained by the intergrowth of 1111-type CaFeAsF and 122-type AFe$_2$As$_2$. While in most IBSs superconductivity emerges after the suppression of antiferromagnetic order in the parent compound by carrier doping, the 12442 compounds are superconducting in their stoichiometric state, with a critical temperature ranging from 28 to 33\,K, without the need for chemical substitution to achieve extra carrier doping. These compounds consist of alternate stacking of conducting Fe$_2$As$_2$ layers and insulating Ca$_2$F$_2$ blocks, and contrary to other IBSs, they have double FeAs layers between neighboring insulating layers, thus mimicking the case of double CuO$_2$ layers in the cuprates. As the cuprates (and contrary to 122 and 1144 IBSs), these compounds show a large anisotropy in the upper critical field and penetration depth\,\cite{Torsello2022NPJQM}. As for the pairing symmetry and gap structure, to date it is still controversial whether in 12442 there exist gap nodes or not, and, in the former case, whether nodal $d$-wave-like or  accidental nodal $s$-wave gaps should be expected.
Recently, a study of the gap structure of RbCa$_2$Fe$_4$As$_4$F$_2$ (Rb-12442) led to the identification of two gaps with signatures of $d$-wave-like nodal behavior\,\cite{Torsello2022NPJQM}.
Data are well described by a two-band $d{-}d$ state with symmetry-imposed nodes, since they persist upon Ni doping.
Within this framework, the need emerges to investigate the role of disorder of nonchemical origin, e.g. induced by ion irradiation. In fact, in this case it is possible to compare measurements on the very same crystals before and after irradiation, i.e. after the introduction of structural defects with defined shape and dimensions, and controlled density. This could help in completing the picture and give hints toward an overall interpretation of the superconductivity mechanism in the 12442 IBSs.

In this work, we report on the effects of disorder introduced via 3.5\,MeV proton irradiation on the critical temperature, superfluid density and gap values of Rb-12442 single crystals, investigated by a combination of the complementary techniques of electric transport measurements, coplanar waveguide resonator (CPWR) measurements and point-contact Andreev-reflection spectroscopy (PCARS).\\
Transport measurements allow the determination of the resistivity as a function of temperature, of the residual resistivity ratio RRR, and of the superconducting transition temperature $T\ped{c}^{\rho}$, and thus provide information on how irradiation affects the scattering of electrons by defects and the onset of the superconducting state.\\
CPWR measurements allow a direct determination of the London penetration depth $\lambda\ped{L}$ as a function of temperature, and consequently of the superfluid density $\rho\ped{s}$ and of the critical temperature $T\ped{c}^{\lambda}$ at which it vanishes. The temperature dependence of $\rho\ped{s}$ provides indications on the symmetry of the superconducting gap(s), and in particular on whether unpaired quasiparticles persist down to the lowest temperatures, which is associated to an incomplete gapping of the Fermi surface typical for example of $d$-wave superconductivity.\\
PCARS allows a direct determination of the amplitude and symmetry of the superconducting gap(s) and of the temperature at which they close, $T\ped{c}\apex{A}$, by measuring the differential conductance $dI/dV$ of a small (point-like) contact between a normal metal (N) and the superconductor under study (S), as a function of the bias voltage $V$ applied across the NS junction\,\cite{Andreev, DagheroSUST2010, DagheroRoPP2011}. The absence of an insulating barrier between the N and S banks enables Andreev reflection at the interface\,\cite{Andreev, DagheroSUST2010, DagheroRoPP2011}, which enhances the $dI/dV$ for $V \leq  \Delta/e$ ($\Delta$ being the superconducting gap) and makes it sensitive to the coherence of the superconducting condensate -- unlike tunnelling spectroscopy that probes the density of states of unpaired quasiparticles. \\
Finally, the dependence of the critical temperature ($T\ped{c}^{\rho}$ or $T\ped{c}^{\lambda}$) on the disorder (which is directly related to the symmetry of the energy gap in single-band systems) can be compared to that observed in other, more widely-studied IBS systems.

The conclusion of this extensive study is that the critical temperature of 12442 decreases on increasing disorder in an unusually weak (if compared to other IBSs) and nonlinear manner, while the London penetration depth increases in an approximately linear way. The joint analysis of PCARS spectra and CPWR measurements supports the existence of nodes in at least one gap, and most probably indicates the persistence of the same gap symmetry observed in the pristine, non-irradiated compound.

\section{Materials and methods}

\subsection{Crystal growth and basic characterization}
Single crystals of Rb-12442 were grown by the self-flux method\,\cite{Wang2019JPCC, Xing2020sust, Yi2020NJP} and their high quality was assessed through X-ray diffraction and energy-dispersive X-ray spectroscopy as detailed in Refs.\,\cite{Yi2020NJP, Torsello2022NPJQM}.

\subsection{Electrical resistivity measurements}
The temperature dependence of the electrical resistivity was measured on selected single crystals by means of the van der Pauw method\,\cite{vanderpauw} after having electrically contacted them with thin gold wires and conducting silver paste. Each resistance measurement was performed by sourcing a small DC current $\approx 100\,\upmu$A between two adjacent contacts and measuring the resulting voltage drop across the opposite pair of contacts. Common-mode offsets including thermoelectric voltages were removed using the current-reversal method.

\subsection{CPWR measurements}
The London penetration depth and superfluid density were measured by means of a coplanar waveguide resonator (CPWR) technique, particularly suitable to study small IBS crystals for irradiation experiments since it is minimally invasive and the same samples can be characterized several times before and after different irradiation steps\,\cite{Ghigo2022Springer2}. 
The measurement was carried out within a resonator perturbation approach. A coplanar waveguide resonator obtained by patterning an YBa$_2$Cu$_3$O$_{7-\delta}$ thin film deposited on MgO substrate was coupled to a vector network analyzer for the measurement of the scattering matrix at low input power. The transmitted power shows a resonance that can be fitted by a modified Lorentzian function, yielding the resonance frequency and quality factor. Measurements were carried out as a function of the temperature for the empty resonator and with the investigated samples coupled to it (\textit{i.e.} placed on the central stripline, far from the edges). The presence of the sample causes shifts of the resonance frequency and of the quality factor that are due to the electromagnetic properties of the sample: from these shifts, after a self-consistent calibration procedure, the absolute value of the penetration depth and its temperature dependence can be assessed\,\cite{Ghigo2022materials}.

\subsection{PCARS measurements}
The structure of the superconducting gap was assessed via point-contact Andreev reflection spectroscopy (PCARS), by measuring the differential conductance ($dI/dV$) of point-like contacts made between a normal metal (N) and the superconducting sample (S).
Unlike in tunnelling spectroscopy, in a direct N/S contact without insulating barrier, electrons impinging on the N/S interface are either transmitted from the metal to the superconductor -- if their energy $eV$ is higher than the gap $\Delta$ in the S side -- or experience Andreev reflection, i.e. are transmitted as a Cooper pair in S while a hole is reflected back in N\,\cite{Andreev, DagheroSUST2010, DagheroRoPP2011}. As a result, the differential conductance of the junction is enhanced for $V$ smaller than $\Delta/e$. In real junctions, a small potential barrier can be present at the interface, which makes a tunneling contribution be present in addition to the Andreev-reflection signal. The shape of the resulting $dI/dV$ vs. $V$ curve depends not only on the gap amplitude, but also  on its symmetry in the reciprocal space. In particular, isotropic gaps give rise to conductance curves with a zero-bias minimum and two symmetric maxima at the gap edges, no matter what is the direction of current injection with respect to the crystallographic axes \cite{DagheroSUST2010}. Instead, anisotropic nodal gaps (especially in the case of a $d$-wave symmetry) can give rise to a variety of different curves depending on the angle of current injection \cite{Daghero_NatComm2012}. For example, zero-bias cusps or peaks are typical features that can be associated to a $d$-wave gap, in specific conditions of barrier height and current direction \cite{DagheroRoPP2011}. In multigap systems, features such as peaks or shoulders are observed in the differential conductance, which are the hallmarks of the different gaps\,\cite{DagheroRoPP2011}.

Point contacts were fabricated by using the so-called soft technique\,\cite{DagheroSUST2010, DagheroRoPP2011}, i.e. by using a thin gold wire ($\diameter \approx 12.7\,\upmu$m) stretched over the platelet-like sample and touching its very thin and flat side surface in a single point. These contacts were (optionally) mechanically stabilized by small ($\diameter \approx 50\,\upmu$m) drop-cast droplets of silver paste. In either case, the ``macroscopic" contact actually consists of several parallel nanoscopic N/S contacts.

The differential conductance was obtained by numerical derivation of the $I-V$ characteristics of the contact, measured in the pseudo-four-probe configuration as detailed in Refs.\,\cite{Piatti2023LTP, Torsello2022NPJQM}. Within the present paper, the direction of (main) current injection was confined along the $ab$ planes of the crystals by making the point contacts on the thin, but smooth and flat side surfaces of the crystals.
For the $dI/dV$ vs. $V$ curves to contain spectroscopic information, the (maximum) energy of the injected electrons must be directly related to the bias voltage applied through the junction, i.e. $E = eV$. This requires that the resistance of the contact is much larger than the resistance of the normal bank, and that electrons do not lose energy while crossing the interface. The latter condition means that there is no Joule dissipation in the contact, and is fulfilled if the individual nanocontacts are in the ballistic or at most diffusive regime \cite{Naidyuklibro,DagheroSUST2010}, i.e. their radius is smaller than the (inelastic) electronic mean free path. 

Information about the gap number, amplitude and symmetry were extracted from  the experimental $dI/dV$ vs. $V$ curves of spectroscopic contacts by fitting them with suitable models.
Before fitting, the experimental curves at any temperature were normalized to the normal-state curve measured just above the superconducting transition -- as customary in IBSs\,\cite{DagheroSUST2010, DagheroPRB2020, DagheroSUST2018} -- after this was rescaled by subtracting the contribution of the so-called spreading resistance $R_s$, i.e. normal-state resistance of the crystal that appears, in series with the contact, in the proximity of the resistive transition and results in a horizontal stretching and a downward shift of the $dI/dV$ spectra with respect to those at low temperature\,\cite{Daghero2014}. As discussed in Refs.\,\cite{Piatti2023LTP, Torsello2022NPJQM}, for each contact we typically selected the value of $R_s$ that allows matching the tails of the normal and superconducting $dI/dV$ spectra. Since this choice remains somewhat arbitrary, we used multiple reasonable values of $R_s$ to obtain the normalized spectra, and repeated each time the fitting procedure. The resulting values of the fit parameters were then averaged and their maximum differences taken as the uncertainty of the procedure. 

The normalized $dI/dV$ spectra were then fitted to the two-band version\,\cite{DagheroSUST2010, DagheroRoPP2011} of the appropriate anisotropic Blonder-Tinkham-Klapwijk model\,\cite{TanakaPRL1995, KashiwayaPRB1995, KashiwayaPRB1996}, as extensively discussed in Refs.\,\cite{Piatti2023LTP, Torsello2022NPJQM}. The reason why we used an effective two-band model even if we expect more than two gaps to be present in this compound is due to the necessity to limit the number of adjustable parameters. Indeed, 
 for each gap $i$ the model includes as adjustable parameters the gap amplitude $\Delta_i$, the broadening parameter $\Gamma_i$, and the barrier parameter $Z_i$. The model also contains the relative weight of the large-gap band in the conductance $w_1$ and the direction $\alpha$ between the direction of (main) current injection and the antinodal line. The latter is crucial in reproducing zero-bias peaks in the experimental spectra arising from the interference effects occurring when electron-like and hole-like quasiparticles experience order parameters of opposite sign. Being based on the weak-coupling BCS theory, the model does not capture strong-coupling features such as electron-boson-derived shoulders appearing at energies higher than the largest gap, which must therefore be excluded from the fitting procedure\,\cite{DagheroRoPP2011, Daghero2014, DagheroPRB2020}.

\subsection{Ion irradiation}
Ion irradiation was carried out at the CN facility of the Legnaro National Laboratories of INFN, at room temperature, in high vacuum, with a defocused 3.5\,MeV proton beam incident parallel to the $c$ axis of the samples. This energy ensures that ion implantation is avoided in all the investigated samples. Beam current was kept below 60 nA on a spot of 7\,mm of diameter, to avoid excessive sample heating.

Irradiation with MeV-protons in IBSs is known to produce point-like defects (Frenkel pairs) and small clusters of nanometric size \cite{Park2018prb}, that in the present case are quite uniformly distributed in the sample, since the Bragg peak is avoided. Such defects are efficient scattering centers that can strongly affect the superconducting state by suppressing the superfluid density.

In order to compare the effects of irradiation on the superconducting properties of our samples with other materials and irradiation conditions we computed the displacements per atom (dpa) through Monte Carlo simulations with the SRIM code\,\cite{Ziegler2010}. This approach has proven useful to study the effects of ion-irradiation-induced disorder because dpa is a good parameter to characterize the amount of scattering centers introduced in the sample\,\cite{Torsello2021SciRep}.
The dpa value is proportional to the proton fluence, that was evaluated through the total deposited charge obtained by integrating the measured beam current at the beam spot on the sample.

\section{Results}

\begin{figure*}[ht!]
\begin{center}
\includegraphics[width=\textwidth]{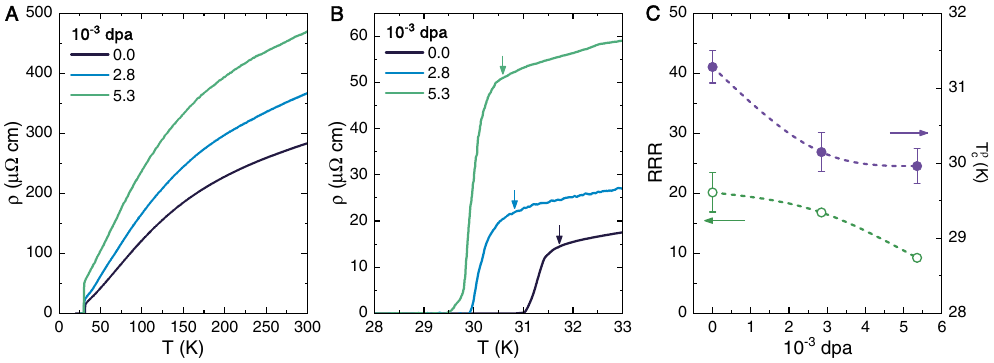}% This is a *.eps file
\end{center}
\caption{
(\textbf{A}) Resistivity $\rho$ as a function of temperature $T$ for three Rb-12442 crystals in the pristine state (corresponding to dpa = 0) and at increasing proton irradiation levels.
(\textbf{B}) Magnification of the curves shown in \textbf{(A)} in the vicinity of the superconducting transition. The arrows highlight the onset temperature of the resistive transition $T\ped{c}\apex{on}$ (see text for details).
(\textbf{C}) Superconducting transition temperature $T\ped{c}\apex{\rho}$ (filled violet circles, right scale) and residual resistivity ratios $RRR$ (hollow green circles, left scale) of the same crystals as a function of the disorder level measured in displacements per atom (dpa). Dashed lines are guides to the eye.}\label{fig:resistivity}
\end{figure*}

We first investigate the effects of ion irradiation on the electric transport properties of the Rb-12442 crystals.
Figure\,\ref{fig:resistivity}A shows the electrical resistivity $\rho$ as a function of temperature $T$ of three crystals at different dpa values, and Figure\,\ref{fig:resistivity}B a magnification close to the superconducting transitions. The normal state of the pristine crystal exhibits the typical behaviour as reported in the literature\,\cite{Yi2020NJP, Torsello2022NPJQM}: the residual resistivity ratio, defined as $RRR = \rho(300\,\mathrm{K})/\rho(T\ped{c}\apex{on})$, is $RRR = 20\pm3$ and larger than the literature values ($12-16$)\,\cite{Yi2020NJP, Torsello2022NPJQM,Yi2023}; whereas the resistivity at room $T$ is $\rho(300\,\mathrm{K}) = 290\pm30\,\upmu\Omega$\,cm and lower than the previously-reported values ($\approx1.1$\,m$\Omega$\,cm)\,\cite{Yi2020NJP}. Both features attest the high crystalline quality of our present samples. 
Here, $T\ped{c}\apex{on}$ is defined as the onset of the superconducting transition, i.e. the temperature at which the resistivity starts to deviate from a linear extrapolation of the normal state (highlighted by the arrows in Figure\,\ref{fig:resistivity}B).
The resistive superconducting transition temperature is defined as $T\ped{c}\apex{\rho} = T\ped{c}^{50} \pm \frac{T\ped{c}^{90}-T\ped{c}^{10}}{2}$, where $T\ped{c}^{\alpha}$ is the temperature at which the resistivity reaches $\alpha\%$ of $\rho(T\ped{c}\apex{on})$, and is equal to $T\ped{c}\apex{\rho} = 31.3\pm0.2$\,K for the pristine crystal, in good agreement with the previous reports\,\cite{Piatti2023LTP, Yi2020NJP, Torsello2022NPJQM}.

At the increase of the irradiation level, the resistivity increases in the entire temperature range up to $\rho(300\,K)=470\pm50\,\upmu\Omega$\,cm in the crystal at $5.3{\times}10^{-3}$\,dpa, and at the same time both $T\ped{c}^\rho$ and $RRR$ are suppressed. As shown in Figure\,\ref{fig:resistivity}C, the suppression of $T\ped{c}^\rho$ introduced by irradiation is significantly faster at lower dpa than at higher dpa values, and for both quantities it is minute with respect to those introduced by substitutional Ni doping. Specifically, we observe a maximum $T\ped{c}\apex{\rho}$ suppression of $4.2\%$ and a minimum $RRR=9$ in our irradiated samples, against a complete disappearance of superconductivity and a minimum $RRR\approx1$ in Ni-doped samples. This indicates that the perturbations to the structural and electronic properties of Rb-12442 crystals introduced by irradiation are minimal with respect to those attained by substitutional doping.

\begin{figure*}[ht!]
\begin{center}
\includegraphics[width=0.85\textwidth]{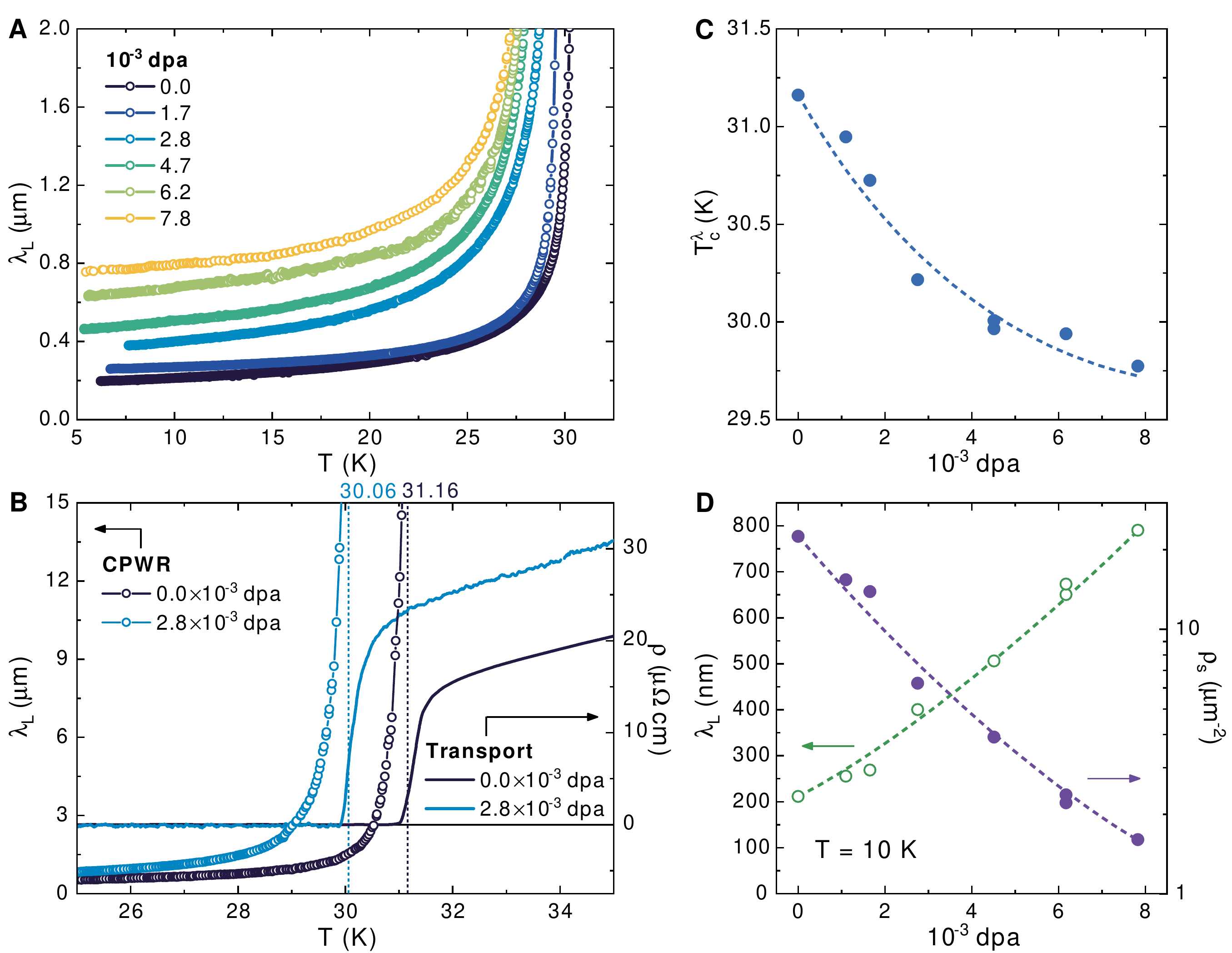}% This is a *.eps file
\end{center}
\caption{
(\textbf{A}) Temperature ($T$) dependence of the London penetration depth $\lambda\ped{L}$ determined from CPWR measurements for pristine and irradiated crystals for increasing proton irradiation levels, measured as displacements per atom (dpa).
(\textbf{B}) Comparison of the $T$ dependencies of $\lambda\ped{L}$ (hollow circles, left scale) and of the electrical resistivity $\rho$ (solid lines, right scale) measured in the same two samples (a pristine crystal and an irradiated crystals at $2.8{\times}10^{-3}\,$dpa). The superconducting critical temperature determined from CPWR measurements, $T\ped{c}^\lambda$, (defined as the $T$ at which $\lambda\ped{L}$ diverges) correlates well with the midpoint of the resistive transition.
(\textbf{C}) $T\ped{c}^\lambda$ and 
(\textbf{D}) $\lambda\ped{L}$ (hollow green circles, left scale) and $\rho\ped{s}$ (filled violet circles, right scales) measured at 10\,K determined from the curves shown in panel (\textbf{A}) as a function of the disorder level measured in dpa. Dashed lines are guides to the eye.}\label{fig:CPWR}
\end{figure*}

A more extensive characterization of the superconducting properties of the Rb-12442 crystals upon increasing ion irradiation was carried out by analyzing the data from CPWR measurements. Figure\,\ref{fig:CPWR}A shows the temperature dependence of the London penetration depth $\lambda\ped{L}$ for the pristine crystal and for five irradiated crystals with increasing dpa values, from 1.7 up to $7.8{\times}10^{-3}$\,dpa. It can clearly be seen that ion irradiation increases $\lambda\ped{L}$ in the entire $T$ range and suppresses the onset temperature below which superconducting behaviour is observed.
In particular, the superconducting critical temperature $T\ped{c}^{\lambda}$ can be obtained as the value of $T$ at which $\lambda\ped{L}$ diverges (or equivalently, at which the superfluid density $\rho\ped{s}=\lambda\ped{L}^{-2}$ goes to 0). Figure\,\ref{fig:CPWR}B shows how this definition correlates extremely well to the midpoint of the resistive transition in transport measurements, for both pristine and irradiated samples.

Figure\,\ref{fig:CPWR}C tracks the evolution of $T\ped{c}^\lambda$ upon increasing irradiation level, which confirms both its small suppression as compared with those attained via substitutional doping (maximum $T\ped{c}^\lambda$ suppression of 4.5\% at $7.8{\times}10^{-3}$\,dpa) and the progressive reduction in its suppression rate at higher dpa values.
Figure\,\ref{fig:CPWR}D shows $\lambda\ped{L}$ measured at $T=10$\,K and the corresponding values of $\rho\ped{s}$ as a function of the dpa values: $\lambda\ped{L}(10\,\mathrm{K})$ increases from 211 to 790\,nm from the pristine crystal to the crystal irradiated at $7.8{\times}10^{-3}$\,dpa, corresponding to a reduction in $\rho\ped{s}(10\,\mathrm{K})$ from 22.4 to 7.8\,$\upmu$m$^{-2}$.
This shows that ion irradiation is much more effective at tuning the magnetic penetration depth of the Rb-12442 crystals with respect to their superconducting transition temperatures, also when compared with Ni substitutional doping: for instance, 5\%\,Ni-doped crystals -- where $T\ped{c}^{\lambda}$ was suppressed by $\approx 34\%$ with respect to the pristine crystals -- exhibited an increase of $\lambda\ped{L}(10\,\mathrm{K})$ only up to $\approx 600$\,nm\,\cite{Torsello2022NPJQM}.

\begin{figure*}[ht!]
\begin{center}
\includegraphics[width=\textwidth]{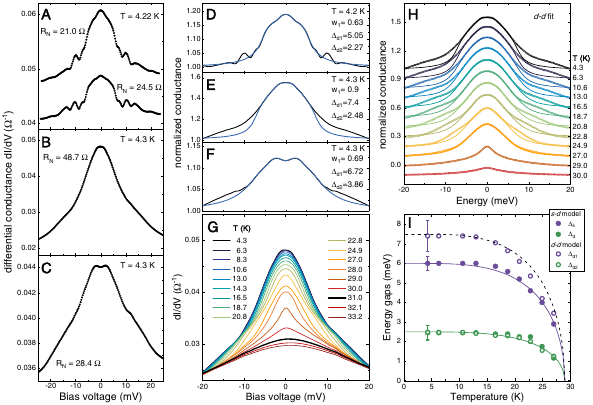} % This is a *.eps file
\end{center}
\caption{
(\textbf{A}),(\textbf{B}),(\textbf{C}) Some examples of as-measured conductance curves ($dI/dV$ vs $V$) of $ab$-plane contacts made on the irradiated crystal for which the disorder level is $5.3{\times}10^{-3}$\,dpa. The normal-state resistance of each contact is indicated in the labels.  (\textbf{D}),(\textbf{E}),(\textbf{F}) Examples of normalized conductance curves (black lines) with the relevant $d{-}d$ fit (blue lines). The labels indicate the temperature at which the spectra were acquired, the value of the weight $w_1$ given by the fit, and the amplitude of the two gaps. (\textbf{G}) Temperature dependence of the raw conductance spectra of the contact with $R_N = 48.7\,\Omega$. The black curve (at $T=31$\,K) is the first that does not show zero-bias features and corresponds to the attainment of the bulk normal state. (\textbf{H}) The same conductance curves of panel (\textbf{G}) after normalization (symbols) and the relevant best-fitting curves (lines) calculated within the $d{-}d$ model. (\textbf{I}) Temperature dependence of the energy gaps extracted from the fit of the curves in panel (\textbf{G}) using the $s{-}d$ model (filled circles) and the $d{-}d$ model (hollow circles). Solid lines are functions of the form of Equation\,\eqref{eq:BCS}.
 }\label{fig:PCARS}
\end{figure*}

Now we turn our attention to the effect of ion irradiation on the values of the superconducting energy gaps as measured by PCARS. Figures\,\ref{fig:PCARS}A, \ref{fig:PCARS}B and \ref{fig:PCARS}C show some examples of low-temperature PCARS spectra of different point contacts (with different normal-state resistance) made on the sample with a disorder level of $5.3{\times}10^{-3}$\,dpa.
Most spectra (Figure\,\ref{fig:PCARS}A) show a zero-bias cusp or maximum, which is suggestive of a nodal gap\,\cite{DagheroSUST2010, DagheroRoPP2011, TanakaPRL1995, KashiwayaPRB1995, KashiwayaPRB1996}, and is perfectly consistent with the observations made on pristine Rb-12442\,\cite{Piatti2023LTP, Torsello2022NPJQM}. Others, instead, show a broad maximum (Figure\,\ref{fig:PCARS}B) or even two symmetric maxima at low bias (Figure\,\ref{fig:PCARS}C) that are usually associated to an isotropic gap, but can be observed even in $d$-wave gap symmetry when the current is injected (mainly) along the antinodal direction\,\cite{DagheroSUST2010, DagheroRoPP2011, TanakaPRL1995, KashiwayaPRB1995, KashiwayaPRB1996}.

More detailed information about the energy gap can be extracted by normalizing the experimental spectra to the normal-state curve measured immediately above $T\ped{c}$\,\cite{DagheroSUST2010, DagheroPRB2020, DagheroSUST2018}, and fitting the normalized spectra to a suitable model for Andreev reflection at a normal metal/superconductor interface\,\cite{DagheroSUST2010, DagheroRoPP2011}. 
Figures\,\ref{fig:PCARS}D, \ref{fig:PCARS}E, and \ref{fig:PCARS}F show three different normalized spectra (black lines). None of these could be fitted by using a single gap, either in $s$-wave or $d$-wave symmetry, indicating that irradiation does not change the multiband nature of superconductivity in this compound. We then used an effective two-band model, in which the relative weight of the gaps is an adjustable parameter. Based on the results found in pristine, unirradiated Rb-12442\,\cite{Piatti2023LTP, Torsello2022NPJQM} and on the fact that some of the curves (Figure \ref{fig:PCARS}D) display a zero-bias cusp, we tried to fit the curves using either an isotropic $s$-wave gap and a nodal $d$-wave gap ($s{-}d$ model in the following) or two $d$-wave gaps ($d{-}d$ model).
As discussed in Ref.\,\cite{Piatti2023LTP}, the $s{-}d$ model actually mimics the nodal $s_\pm$ and the $d_{x^2-y^2}$ symmetries, whereas the $d{-}d$ model mimics the $d_{xy}$ symmetry, all compatible with the tetragonal structure of Rb-12442\,\cite{Hirschfeld}. In both the models, the partial conductance associated to each gap is modelled as in Ref.\,\cite{KashiwayaPRB1996}. The fitting parameters are then, in addition to the gap amplitudes $\Delta_s$ and $\Delta_d$ (in the $s{-}d$ model) or $\Delta_{d1}$ and $\Delta_{d2}$ (in the $d{-}d$ model), the weight $w_1$ of the large gap (such that $w_2=1-w_1$), the barrier parameters $Z_1$ and $Z_2$, the broadening parameters $\Gamma_1$ and $\Gamma_2$, and the angle $\alpha$ between the direction of (main) current injection and the antinodal direction of the $d$-wave gap. In the case of the $d{-}d$ model, we assume $\alpha_1 = \alpha_2$, that implies that both the gaps have nodes in the same direction.

It turns out that both the models allow a good fit of the curves, as already pointed out in the case of pristine Rb-12442 and of the Ni-doped compound\,\cite{Piatti2023LTP, Torsello2022NPJQM}. The fact that irradiation does not change the gap structure is compatible both with the small effect on the critical temperature, and with the observed persistence of the same structure even in Ni-doped samples\,\cite{Piatti2023LTP, Torsello2022NPJQM} where the dopants introduce disorder and give rise to a much more significant $T\ped{c}$ suppression (about 10\,K). It is worth noting that, in order to reproduce the shape of the curves in panels E and F, the current injection has to be assumed to occur along the antinodal direction of the $d$-wave gap (i.e. $\alpha\simeq0$).
Just as an example, Figures\,\ref{fig:PCARS}D, \ref{fig:PCARS}E and \ref{fig:PCARS}F show the results of the $d{-}d$ fit of three different curves measured at the lowest temperature. The fit is very good in the central part of the curves, while for higher energies (starting at about 10\,meV) it fails due to an excess conductance probably due to the coupling of charge carriers with some bosonic mode involved in the superconducting pairing, as always observed in IBSs\,\cite{Tortello2010, DagheroRoPP2011, Daghero2014, DagheroPRB2020}.
The amplitudes of the superconducting gaps vary from one contact to another, probably due to the fact that the compound features more than two gaps, while we are using here a minimal (effective) two-gap model in order to keep the number of adjustable parameters to a minimum; moreover, the presence of the bosonic structures prevents an accurate determination of the larger gap, presumably causing its overestimation.
The same happened in pristine Rb-12442, where the gap amplitudes were spread over a rather wide band of energies. In particular we found, for the $d{-}d$ model, $\Delta_{d1} \in [4.8, 7.3]$\,meV and $\Delta_{d2} \in [1.3, 3.0]$\,meV. The gap amplitudes that result from the $d{-}d$ fit of the curves in irradiated Rb-12442 are still generally compatible with the same distribution of gap values, meaning that no significant change in the gap amplitudes has been produced by irradiation. The same happens in the case of the $s{-}d$ fit.

Figure\,\ref{fig:PCARS}G reports, as an example, the temperature dependence of the conductance curves of a point-contact whose normal-state resistance is 48.7\,$\Omega$. Above 31\,K (black curve) the conductance curves are smooth and do not show any feature at zero bias. This means that, at this temperature, the transition to the normal state is complete (in accordance with the green curve in Figure\,\ref{fig:resistivity}B). We thus chose this curve as the normal-state conductance and used it to normalize all the others at lower temperatures, with no need of any further rescaling or shift. The resulting normalized curves are shown in Figure\,\ref{fig:PCARS}H (symbols), vertically offset for clarity (apart from the lowest-temperature curve). The best-fitting curves, obtained within the $d{-}d$ model, are shown as well as solid lines.
The fit follows rather well the temperature evolution of the experimental curves; up to 23\,K, it was possible to fit them by keeping all the parameters of the model (apart from the gaps) fixed to their lowest-temperature values, which is always a good indication of both the reliability of the curves (and of the normalization) and of the ability of the model to grasp the physics of the system. In particular, we used $w=0.9$, $\Gamma_1 = 0.45$\,meV, $Z_1=0.14$, $\Gamma_2 =0.25$\,meV, $Z_2=0.32$. Above 23\,K, instead, the conductance curves start to shrink, becoming too narrow with respect to what predicted by the model; if one wants to reproduce this shrinking, the values of the barrier parameters and of the broadening parameters must be reduced.
The possible reason for this behaviour can be understood by plotting the gap amplitudes resulting from the fit as a function of the temperature, as in Figure\,\ref{fig:PCARS}I. This plot shows that the small gap amplitude is rather robust against the model used (either $s{-}d$ or $d{-}d$) which is natural since in either case the small gap has the $d$-wave symmetry. The amplitude of the larger gap instead depends on whether it is $s$-wave or $d$-wave, which is perfectly natural due to the different shape of the Andreev-reflection features that these two symmetries produce. The most relevant thing is that, no matter what the model is, the gap amplitudes follow a very clear trend which can be approximated by the solid and dashed curves, whose expression is
\begin{equation}
    \Delta_i(T) = \Delta_i(0)\tanh \left(2.0 \sqrt{ \frac{T\ped{c}\apex{A}}{T}-1}\right)
    \label{eq:BCS}
\end{equation}
where $\Delta_i$ is the gap of the $i$-th band and $T\ped{c}\apex{A} =29$\,K. 
This indicates that the behaviour is not exactly BCS, and that the Andreev critical temperature $T\ped{c}\apex{A}$ (i.e. the temperature at which the Andreev signal disappears, and at which the superconducting bank of the junction becomes normal) is about 29\,K, and thus smaller than $T\ped{c}=31$\,K at which the narrow zero-bias maximum in the experimental curves disappears (see Figure\,\ref{fig:PCARS}G). As is clear from Figure\,\ref{fig:resistivity}B, $T\ped{c}\apex{A}$ falls in the proximity of the superconducting transition, but still in a region where the resistivity is zero.
\begin{figure}[ht!]
\begin{center}
\includegraphics[width=\columnwidth]{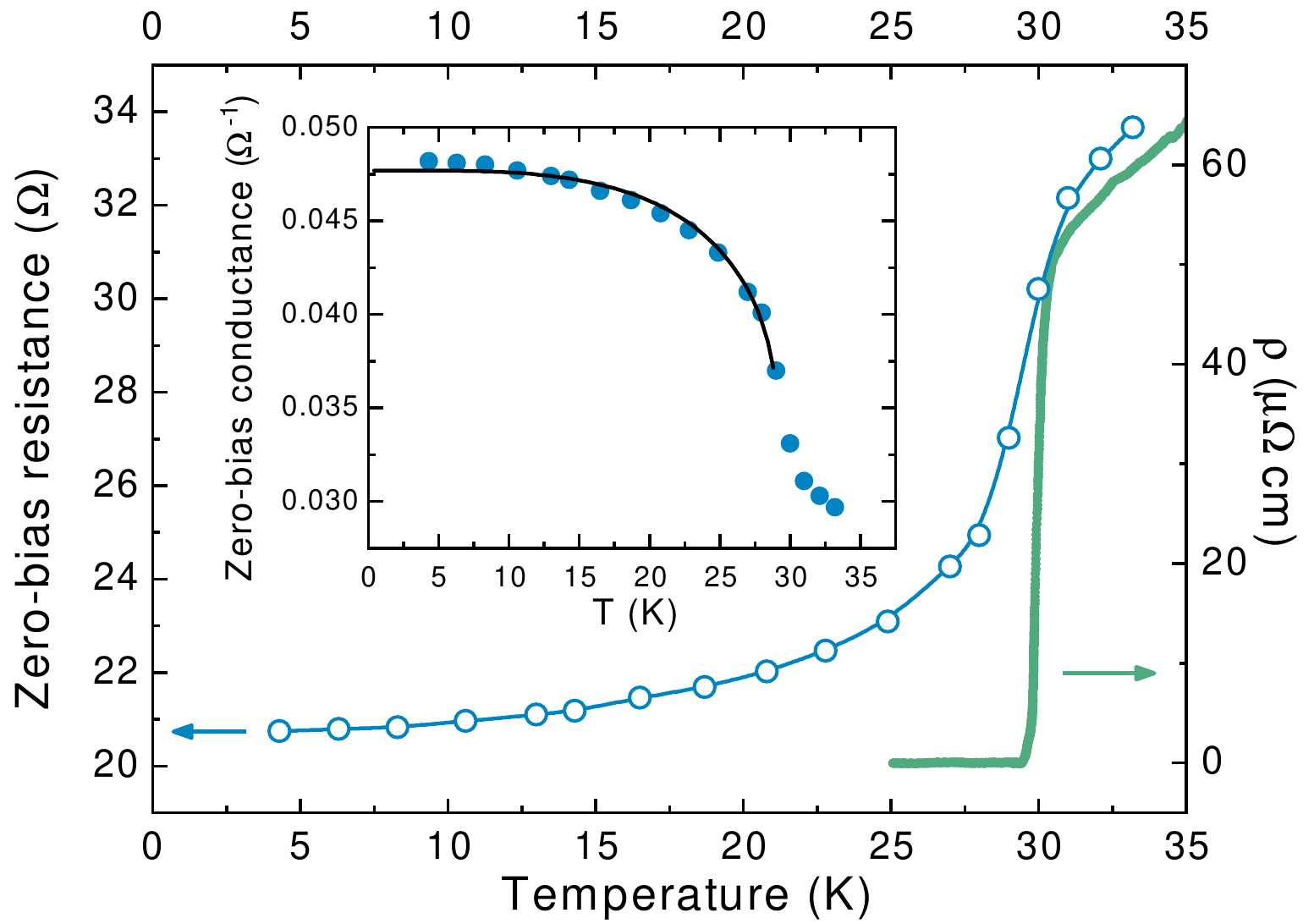} % This is a *.eps file
\end{center}
\caption{
Temperature dependence of the zero-bias resistance extracted from the curves in Figure\,\ref{fig:PCARS}G (blue hollow circles, left scale) compared to the resistivity of the same crystal (solid green line, right scale). The inset shows the zero-bias conductance (filled blue circles), that follows the same temperature dependence of the gaps up to 29\,K (solid black line).}\label{fig:zero_bias}
\end{figure}
This is likely to be due to the non perfectly ballistic nature of the contact, so that a very small heating occurs in the contact region giving rise to a small mismatch between the “bulk" $T\ped{c}^\rho$ and $T\ped{c}\apex{A}$. The same effect explains the narrow zero-bias peak that emerges at high temperature in the conductance curves: while at zero bias the temperature of the contact is equal to that of the bulk, on increasing the bias voltage the current gives rise to heating and, being close to the normal transition, this produces an increase in resistance (and thus a decrease in conductance). Such an effect ceases when the temperature at which the curve is acquired is already in a region where the normal transition (as measured by transport) is complete, since here the resistivity has a much weaker dependence on temperature. Figure\,\ref{fig:zero_bias} displays the temperature dependence of the zero-bias resistance (main panel) and conductance (inset). It is clear that the high-temperature behaviour of both these quantities is no longer due to the temperature dependence of the superconducting gap, but is related instead to the temperature dependence of the crystal's resistivity.

\section{Discussion}\label{sec:discussion}

\begin{figure*}[ht!]
\begin{center}
\includegraphics[width=0.85\textwidth]{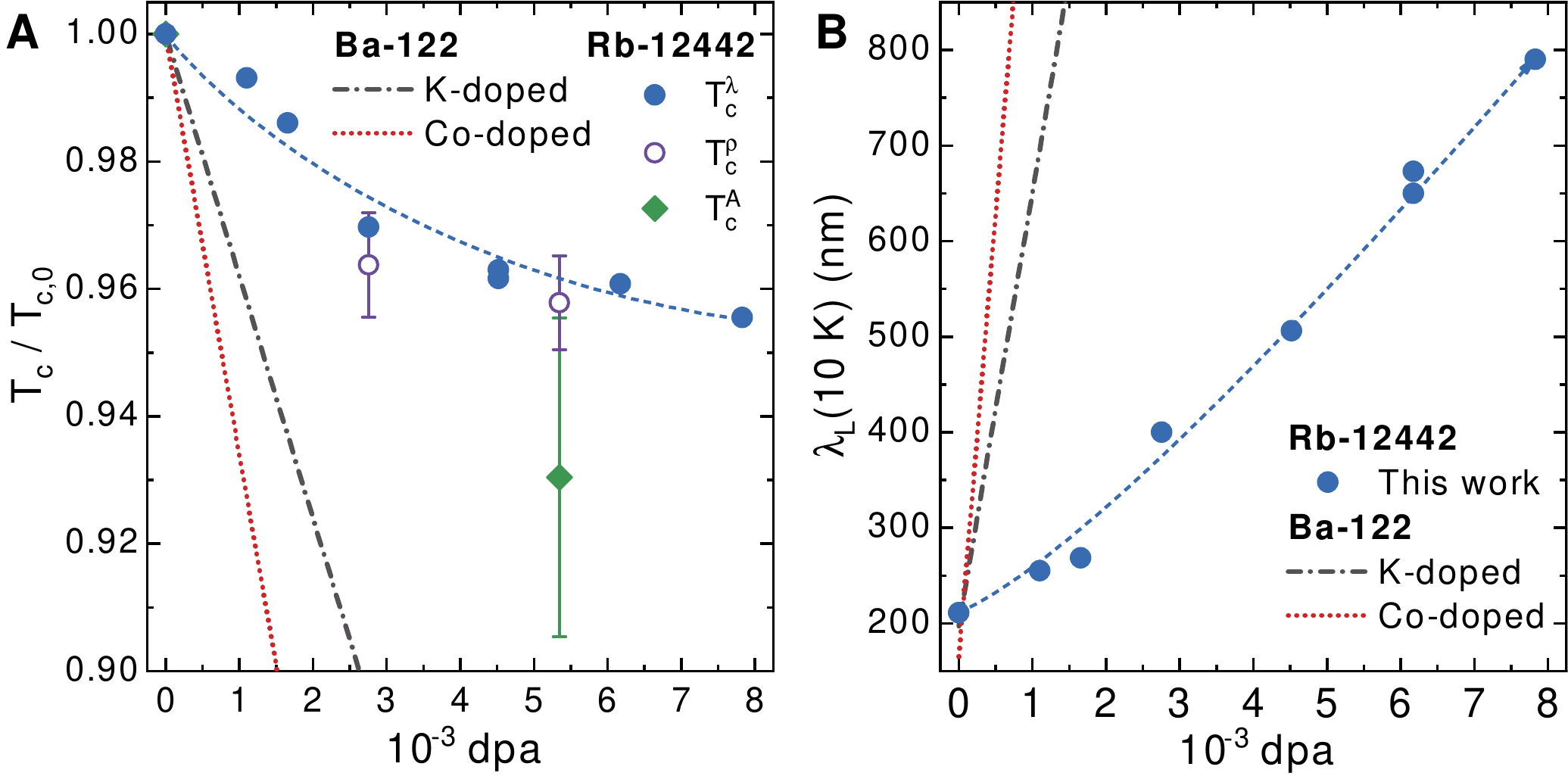}% This is a *.eps file
\end{center}
\caption{(\textbf{A}) Proton-irradiation-induced suppression of the superconducting critical temperature $T\ped{c}$ for Rb-12442 samples, determined from CPWR analysis ($T\ped{c}^\lambda$, filled blue circles), from dc resistivity ($T\ped{c}^\rho$, hollow violet circles) and from PCARS measurements ($T\ped{c}\apex{A}$, filled green diamonds -- the smaller value is due to localized contact heating as explained in the text). The dashed line is a guide for the eye. Data are normalized to the corresponding transition temperature at 0\,dpa ($T\ped{c,0}$) and compared to the trend (black and red lines) showed by two optimally-doped systems of the 122 IBS family (K and Co-doped BaFe$_2$As$_2$), for which the scaling laws with dpa were reported\,\cite{Torsello2021SciRep}. (\textbf{B}) Proton irradiation-induced increase of the London penetration depth $\lambda\ped{L}$ at $T=10$\,K for Rb-12442 samples (symbols, dashed line is a guide to the eye), compared to the trends shown by the same two IBS Ba-122 systems shown in (\textbf{A})\,\cite{Torsello2021SciRep}.}\label{fig:Tcdpa}
\end{figure*}

The response to proton-irradiation-induced disorder of Rb-12442 is qualitatively as expected for the IBSs, showing a suppression of the critical temperature and of the superfluid density due to the enhanced scattering of carriers in a multiband system. However, there is a relevant quantitative difference: below about $4-5{\times}10^{-3}$\,dpa, the $T\ped{c}$ suppression rate with dpa in optimally-doped (Ba,K)Fe$_2$As$_2$ is about twice than for Rb-12442 samples, despite the similar pristine critical temperature of the two compositions. This is reported in Figure\,\ref{fig:Tcdpa}A, where the data obtained in this work on Rb-12442 are compared to the typical behavior exhibited by the 122 family of IBSs when exposed to the same irradiation process\,\cite{Torsello2021SciRep}.
It is also evident that the $T\ped{c}$ vs dpa curve of Rb-12442 shows an upward curvature (or alternatively, that it can be divided in two separate linear trends with different slopes), implying an even lower dpa dependence at higher disorder levels. 
This behavior is quite unexpected, and could -- in principle -- be the result of an important change in the symmetry of the superconducting gap of the system. However, the fact that the experimental PCARS spectra measured on a sample with a disorder level of $5.3{\times}10^{-3}$\,dpa can be satisfactorily fitted using models employing the same gap symmetry describing pristine and Ni-doped Rb-12442\,\cite{Piatti2023LTP, Torsello2022NPJQM} makes this scenario highly unlikely.
Similarly to $T\ped{c}$, also the low-temperature value of the London penetration depth, as shown in Figure\,\ref{fig:Tcdpa}B, has a much smaller variation with dpa when compared to other IBSs. In this case, the modification with dpa is approximately linear in the whole investigated range.

These two observations indicate that the condensate in this system is affected much less than similar systems by the exposition to the same proton irradiation conditions. Since the composition, density and chemical nature of the 12442 and 122 families of IBSs is similar, a similar defects morphology is induced in the samples. Therefore, either the recombination rate of the created defects is much higher in the 12442 family (possibly due to the more layered and anisotropic structure), or the superconducting state itself is more radiation-hard.
The latter option could be related to the other peculiarities observed in this system, such as the observation of nodal behavior\,\cite{Torsello2022NPJQM, Smidman2018PRB, Kirschner2018prb, Wang2020SciChi} and its vicinity to the disappearance of the electron-like pockets \cite{Piatti2023LTP}.

\begin{figure*}[ht!]
\begin{center}
\includegraphics[width=0.85\textwidth]{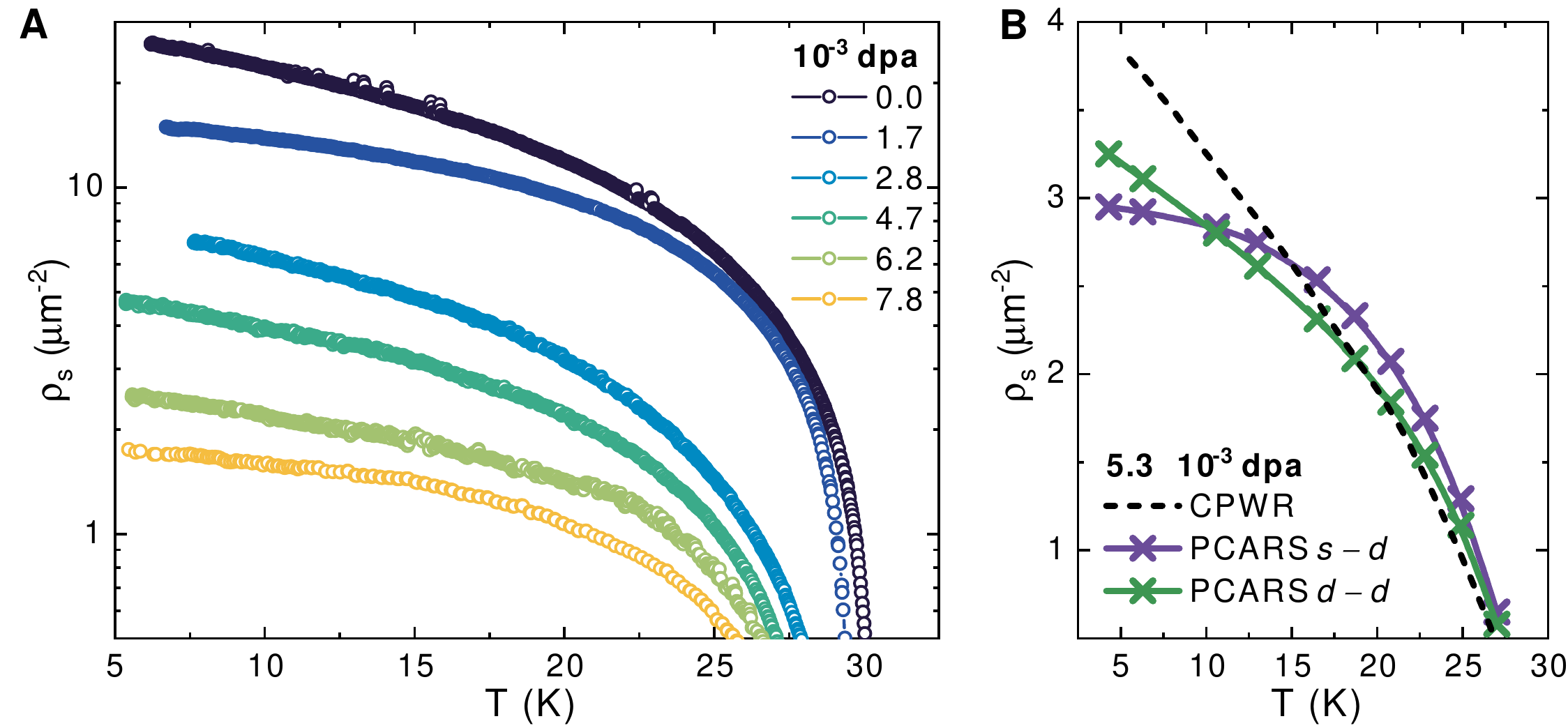}% This is a *.eps file
\end{center}
\caption{
(\textbf{A}) Superfluid density ($\rho\ped{s}=\lambda\ped{L}^{-2}$) as a function of temperature $T$ for increasing values of proton irradiation-induced disorder measured in displacements per atom (dpa), from CPWR measurements.
(\textbf{B}) Data calculated from the gap values measured by PCARS (crosses) are compared to the temperature dependence of $\rho\ped{s}$ deduced starting from CPWR data for the disorder level of $5.3{\times}10^{-3}$\,dpa.
}\label{fig:rho}
\end{figure*}

A further confirmation of the robustness of nodal behavior in Rb-12442 against the introduction of disorder via proton irradiation can be obtained from the temperature dependence of the superfluid density. Figure\,\ref{fig:rho}A shows $\rho\ped{s}=\lambda\ped{L}^{-2}$ as a function of $T$ determined from the $\lambda\ped{L}(T)$ curves shown in Figure\,\ref{fig:CPWR}A: beyond the suppression of the superfluid density in the entire temperature range with increasing dpa, it can be seen that all the $\rho\ped{s}(T)$ curves exhibit a smooth, kink-less behavior in the experimentally-accessible temperature range, and a clear linear dependence at low temperatures. Both features are consistent with the previous reports in pristine and Ni-doped Rb-12442 crystals\,\cite{Piatti2023LTP, Torsello2022NPJQM} and are typically associated with multigap superconductivity with mainly interband coupling\,\cite{Torsello2019PRB} and with the existence of at least one nodal gap\,\cite{Smidman2018PRB, Kirschner2018prb} respectively.

This interpretation can be further corroborated by combining the results obtained independently by means of the CPWR and PCARS measurements.
As a matter of fact, the superfluid density can also be computed from the zero-temperature London penetration depth, $\lambda\ped{L}(0)$, and from the temperature-dependence of the gap values extracted from PCARS, $\Delta_i(T)$, as\,\cite{Chandrasekhar1993}:
\begin{widetext}
\begin{equation}
    \rho\ped{s}(T)=\frac{1}{\lambda_L(0)^2}\sum_i w_i \left[1+\frac{1}{\uppi}\int_0^{2\uppi}\int_{\Delta_i(\phi,T)}^{\infty} \frac{\partial f}{\partial E}\frac{E dE d\phi}{\sqrt{E^2-\Delta_i^2(\phi,T)}}\right]
\end{equation}
\end{widetext}
where $i$ identifies the band, $\Delta_i(\phi,T)$ is the angle-dependent superconducting gap function, $f = [1 + \exp(E/k_\mathrm{B} T )]^{-1}$ is the Fermi function and $w_i$ is the mixing weight of the $i$-th gap contribution (constrained by $w_1+w_2=1$). For this latter parameter we use the same value previously employed for pristine undoped Rb-12442\,\cite{Torsello2022NPJQM} and consider it as a fixed parameter.
One can factorize the angular and temperature dependencies of the gap as $\Delta_i(\phi,T) =\Delta_{0,i} f_\phi(\phi) f_T(T)$, where the angular function for $d$-wave (nodal) superconductors is $f_\phi(\phi)=\cos(2\phi)$ and the gap amplitude is the one measured by PCARS. To obtain the $\lambda\ped{L}(0)$ value for the disorder level of the sample measured by PCARS we linearly interpolate the data shown in Figure\,\ref{fig:Tcdpa}B.

The comparison between the superfluid data from PCARS obtained within this approach and the data from the CPWR technique is shown in Figure\,\ref{fig:rho}B, for the disorder level of $5.3{\times}10^{-3}$\,dpa. 
Violet (green) crosses indicate the values of $\rho\ped{s}$ calculated by starting from the PCARS gaps extracted from the $s{-}d$ ($d{-}d$) fit, while the black dotted line represents the $\rho_s(T)$ curve deduced from the CPWR analysis for a disorder level of $5.3{\times}10^{-3}$\,dpa. The PCARS $d{-}d$ case seems to yield the best qualitative match with the CPWR trend since it reproduces the linear behavior at low temperatures (even though with a lower slope), contrarily to what is found for the  $s{-}d$ case, where the $s$-wave component forces the curve to saturate.

In summary, we reported on the unusually weak irradiation effects on the superconducting properties of the IBS Rb-12442. The analysis was performed by CPWR and PCARS on single crystals before and after 3.5-MeV proton irradiation, at different damage levels up to $7.8{\times}10^{-3}$\,dpa. Both the critical temperature and the superfluid density are shown to decrease with disorder, but at a much lower rate than for other IBS systems with similar pristine values of $T\ped{c}$ and $\rho\ped{s}$. Moreover, the critical temperature degradation shows also a qualitatively different trend with respect to the other mentioned IBSs, i.e. a change of linear slope or possibly an upward curvature, implying an even smaller disorder dependence for higher irradiation doses. A comparison between results from the two measurement techniques allowed us to elaborate about the order parameter symmetry, in an effective two-gap approach. The evidence of a nodal behavior is confirmed also for the irradiated crystal and on the basis of the comparison we are inclined to select the $d{-}d$ scenario rather than the $s{-}d$ one, even though an ultimate evidence is still needed, which encourages further investigation.

\section*{Conflict of Interest Statement}
The authors declare that the research was conducted in the absence of any commercial or financial relationships that could be construed as a potential conflict of interest.

\section*{Author Contributions}
The work was conceived by D.T. and G.G., who also conducted CPWR measurements and analyzed the results.
E.P. and D.D. performed the electric transport and PCARS measurements and analyzed the results.
M.F., R.G. and L.G. conducted sample irradiation.
X.Y., X.X., and Z.S. grew the crystals.
The manuscript was written by D.T., G.G., E.P. and D.D. with comments and input from all authors. 

\section*{Funding}
This work was supported by the Italian Ministry of Education, University, and Research through the PRIN-2017 program: D.T., M.F., R.G., L.G. and G.G. acknowledge support from project “HIBiSCUS”, Grant No. 201785KWLE; E.P. and D.D. acknowledge support from project “Quantum2D”, Grant No. 2017Z8TS5B). D.T. also acknowledges that this study was carried out within the Ministerial Decree no. 1062/2021 and received funding from the FSE REACT-EU - PON Ricerca e Innovazione 2014-2020. This manuscript reflects only the authors’ views and opinions, neither the European Union nor the European Commission can be considered responsible for them.

\section*{Acknowledgments}
Support by INFN CSN5 and by the European Cooperation in Science and Technology (COST) action CA21144 “SUPERQUMAP” is also acknowledged.

\section*{Data Availability Statement}
The datasets used and analysed for this study are available from the corresponding author on reasonable request.

\bibliography{Torsello_arXiv}

\end{document}